\documentclass[aps,prd,twocolumn,nofootinbib]{revtex4-1}  
\usepackage{graphicx}  
\usepackage{dcolumn}   
\usepackage{bm}        
\usepackage{amssymb}   
\usepackage{xcolor}
\usepackage[normalem]{ulem}
\usepackage{cancel}
\usepackage{amsmath}
\usepackage[english]{babel}

\newcommand{\be}{\begin{equation}}
\newcommand{\ee}{\end{equation}}

\newcommand{\e}{\mbox{e}}
\newcommand{\de}{\mbox{d}}

\begin{document}

\begin{flushleft}
KCL-PH-TH/2016-32
\end{flushleft}

\date{}
\author{Marco de Cesare, Andreas G. A. Pithis, Mairi Sakellariadou}

\affiliation{Department of Physics, King's College London, University
  of London, Strand WC2R 2LS, London, United Kingdom}

\email{marco.de\_cesare@kcl.ac.uk}
\email{andreas.pithis@kcl.ac.uk}
\email{mairi.sakellariadou@kcl.ac.uk}

\title{Cosmological implications of interacting Group Field Theory
  models: cyclic Universe and accelerated expansion}

\begin{abstract}
We study the cosmological implications of interactions between
\emph{spacetime quanta} in the Group Field Theory (GFT) approach to
Quantum Gravity from a phenomenological perspective. Our work
represents a first step towards understanding Early Universe Cosmology
by studying the dynamics of the emergent continuum spacetime, as
obtained from a fundamentally discrete microscopic theory. In
particular, we show how GFT interactions lead to a recollapse of the
Universe while preserving the bounce replacing the initial
singularity, which has already been shown to occur in the free
case. It is remarkable that cyclic cosmologies are thus obtained in
this framework without any \emph{a priori} assumption on the geometry
of spatial sections of the emergent spacetime. Furthermore, we show
how interactions make it possible to have an early epoch of
accelerated expansion, which can be made to last for an arbitrarily
large number of e-folds, without the need to introduce an \emph{ad
  hoc} potential for the scalar field.
\end{abstract}

\maketitle

\section{Introduction}
Our current understanding of cosmological data is based on classical
models that rely on the validity of Einstein's theory of
gravity. However fruitful this approach has been so far, the basic
assumptions on which it relies are clearly unjustified from a
fundamental point of view, and evidence of their violation is expected
to become manifest at the earliest stages of expansion of our
Universe. This is in fact where the dynamics of spacetime should more
appropriately be given in the framework of a Quantum Gravity theory
and a description of spacetime as a continuum medium should make way
for its understanding as fundamentally discrete. Of particular
relevance in this sense is the status of the inflationary paradigm. In
fact, while its success in providing an explanation for structure
growth and solving cosmological puzzles is undeniable, its \emph{ad
  hoc} underlying assumptions do not find support in a more
fundamental theory. More specifically, since the onset of inflation is
supposed to take place at Planckian times, the dynamics of the
Universe at this stage should find a more suitable formulation so as
to take quantum gravitational effects into account. Furthermore, it is
conceivable that the quantum dynamics of the gravitational field
itself could effectively give rise to dynamical features similar to
those of inflationary models, without the need to introduce a new
hypothetical field (the inflaton) with an \emph{ad hoc} potential.

The purpose of this article is to bridge the gap between the quantum
gravity era and the standard classical cosmological model, following
the line of research started in
\cite{GFCEmergentFriedmann,FreeGFTpheno}. As in those works, we adopt
the Group Field Theory approach to Quantum Gravity \cite{GFT} which provides a formal and complete definition of spin foam models \cite{SF} themselves giving a path integral formulation for Loop Quantum Gravity (LQG) \cite{LQG, LQGSF}. Furthermore, GFT provides a second quantized Fock space reformulation of the kinematical Hilbert space of LQG \cite{GFTLQG} which also allows to explore quantum geometries encoded by spin network states with many nodes. In the GFT approach spacetime geometry is seen as emerging from the collective behavior of basic building blocks (also called ``quanta of geometry'') when taking the macroscopic limit. Such quanta represent the fundamental degrees
of freedom of the gravitational field and can be described as
simplices equipped with data of group theoretical nature. They can be
glued together through their interactions to form simplicial
complexes, which in turn correspond to quantum states of spacetime
geometry. GFT is completely background independent and
General Relativity is expected to be recovered dynamically by taking
the macroscopic and continuum limit. The particular case of a
homogeneous and isotropic Universe, which is the one relevant for our
applications, can be obtained by studying the dynamics of an isotropic
condensate of GFT quanta for a large number of constituents. Exploiting the structure of the kinematical Hilbert space of LQG and in particular the discreteness of the spectra of the geometric operators defined thereon \cite{LQGdiscreteness}, the GFT formalism allows for a description of the effective dynamics of the emergent spacetime by means of classical evolution equations for the
expectation values of geometric observables. These can be recast in a
form which is close to the classical Friedmann equation.

The object of our study is the effect of interactions between the
fundamental building blocks, as given in the microscopic theory, on
the dynamics of the emergent spacetime. It has already been shown in previous works~\cite{GFCEmergentFriedmann,FreeGFTpheno} that GFT
predicts the occurrence of a bounce that replaces the initial
spacetime singularity bedeviling classical cosmological
models. In this sense, the results of GFT condensate cosmology are reminiscent of those found in Loop Quantum Cosmology (LQC) where the resolution of the initial singularity was shown to be a robust feature \cite{BojowaldBounce,LQC}. 

The analysis done in Refs.~\cite{GFCEmergentFriedmann,FreeGFTpheno}, however, did not take into account the effect played by GFT interactions. These encode the connectivity of the above
mentioned graph structures, in other words they are responsible for
glueing the fundamental building blocks. Here we show, under fairly
general assumptions, that the same result holds also in the
interacting case. Furthermore, we show that interactions play a
substantial r{\^o}le in determining the ultimate fate of the Universe
and lead to its recollapse. These two results together imply a cyclic
evolution for the Universe. It is remarkable that this prediction has
been obtained without making any a priori assumptions on the geometry
of the spatial slices of the emergent spacetime.

In previous work by some of the authors~\cite{FreeGFTpheno} it has
also been shown that the bounce is accompanied by an early era of
accelerated expansion. However, here we show that its duration in the
free case is subject to very stringent bounds and cannot accommodate
for a reasonably large number of e-folds. Considering a model with an
effective potential which is reminiscent of multicritical models
\cite{SFTQFT}, here we show that it is possible to achieve an
arbitrarily large number of e-folds by imposing a hierarchy between
two interaction coefficients. At the same time, this allows to select
a sub-class of models that realize such dynamics, which might give
indications to rule out certain models as candidates for the
fundamental microscopic theory. It should be stressed that the result
follows only from assuming a particular form of the interactions
between basic building blocks. A minimally coupled massless scalar
field is introduced merely for the purpose of defining a relational
clock and by no means must be identified with the inflaton.

The plan of the paper is as follows. In Section~\ref{sec:Basics} we
review the effective dynamics of GFT condensate and show how the
evolution of macroscopic quantities is extracted from it. We introduce
our model and discuss how a sign ambiguity in the coefficient of the
kinetic term is fixed by means of phenomenological considerations on
the rate of expansion of the Universe at late (relational) times. In
Section~\ref{sec:Recollapse} we show how the higher power interaction
term induces a recollapse and we discuss the cyclicity of solutions of
the model. Section~\ref{sec:GeometricInflation} is divided in two
parts. In the first part, Section~\ref{sec:FreeCase}, we discuss the
case of a non interacting GFT model and show that it does not support
a reasonably large number of e-folds. In the second one,
Section~\ref{sec:MultiCriticalCase} we show how this is made possible
by considering suitable interactions terms. In
Section~\ref{sec:EffectiveFriedmannEffFluids}, we discuss how the
different terms in the GFT effective potential can be reinterpreted,
from the point of view of an effective Friedmann equation, as sources
corresponding to effective fluids with peculiar equations of
state. Finally, in Section~\ref{sec:Outlook}, we review the main
results of the article and conclude discussing possible lines for
future investigations. In the Appendix we show how to make contact
between the relational dynamics employed here and the standard
formulation of FLRW cosmology.

\section{Non linear dynamics of a GFT condensate}\label{sec:Basics}

The dynamics of an isotropic GFT condensate can be described by means
of the effective action \cite{GFCEmergentFriedmann}
\be\label{eq:EffectiveAction}
S=\int\de\phi\;\left(A~|\partial_{\phi}\sigma|^2+\mathcal{V}(\sigma)\right).
\ee Here $\sigma$ is a complex scalar field representing the
configuration of the Bose condensate of GFT quanta as a function of
relational time $\phi$. The form of the effective potential
$\mathcal{V}(\sigma)$ can be motivated by means of the microscopic GFT
model and we require it to be bounded from below.  There is an
ambiguity in the choice of the sign of $A$, which is not fixed by the
microscopic theory and will turn out to be particularly relevant for
the cosmological applications of the model\footnote{This ambiguity has also been discussed earlier in
  Ref.~\cite{GFCCalcagni} when exploring the possibility to embed LQC
  in GFT.}. In particular, it can be
used to restrict the class of microscopic models by selecting only
those that are phenomenologically viable.  In fact, as we will show,
only models entailing $A<0$ are sensible from a phenomenological point
of view since otherwise one would have faster than exponential
expansion.

The dynamics of macroscopic quantities is obtained by computing
expectation values on the condensate of the corresponding physical
observables in the quantum theory. Introducing some notation is now in
order. We define $\rho_j$ as the modulus of the component of the field
$\sigma$ corresponding to the spin-$j$ representation of SU(2),
thus introducing its polar form
\be\label{eq:PolarForm}
\sigma_j=\rho_j\;\e^{i\theta_j}.
\ee
The quantity $V_j\sim \ell_{p}^3 j^{3/2}$ represents an elementary volume
determined by the particular SU(2)-representation adopted. Since the
volume operator is diagonal in the basis of spin representations, one
has for its expectation value 
\be\label{eq:VolumeDefinitionGeneral}
V(\phi)=\sum_{j\in\mathbb{N}_0/2}V_j \rho_j^2(\phi), \ee 
as in Refs.~\cite{GFTOperators, GFCEmergentFriedmann} where it is
shown how geometric operators in GFT are constructed from their first
quantized counterpart in Loop Quantum Gravity (LQG) \cite{LQGdiscreteness, LQG}. The importance of
such operators is paramount since they allow to give a geometric
interpretation of the solutions of the theory, \emph{e.g.}, in the
condensate phase we consider here.
Equation~(\ref{eq:VolumeDefinitionGeneral}) has important applications
to homogeneous and isotropic cosmologies. Indeed, it makes it possible
to study the evolution of the Universe by means of classical evolution
equations, therefore establishing a link with the dynamics of
classical FLRW models.  The evolution of the volume over (relational)
time $\phi$ is determined from the evolution of all the
$\rho_j$'s. Furthermore, it is reasonable to expect that the
condensate field peaks on a particular representation $j$, in which
case the r.h.s. of Eq.~(\ref{eq:VolumeDefinitionGeneral}) would only
consist of one term. In this case we will omit the index $j$ in
$\rho_j$ and write
\be\label{eq:VolumeDefinition} V(\phi)=V_j
\rho^2(\phi).  \ee 
Indeed, such a natural assumption is supported by recent findings in
the same setting used here, which explicitly show that GFT condensates
form a low spin phase of many quanta of geometry which
are almost entirely characterized by only one spin j (the occupation
numbers corresponding to other reprentations are strongly suppressed),
as shown in Ref.~\cite{GFClowspin}. Thus, throughout the article we
will work under the assumption that the spin representation $j$ is
fixed.

The macroscopic dynamics following from the effective GFT action can
be given in terms of effective Friedmann equations, giving the
relational evolution of the volume with respect to the scalar field
$\phi$. Those are obtained by differentiating
Eq.~(\ref{eq:VolumeDefinition})
\begin{align}
\frac{\partial_{\phi}V}{V}&=2\frac{\partial_{\phi}\rho}{\rho},\label{eq:EmergingFriedmannI}\\ \frac{\partial^2_{\phi}V}{V}&=2\left[\frac{\partial^2_{\phi}\rho}{\rho}+\left(\frac{\partial_{\phi}\rho}{\rho}\right)^2\right]\label{eq:EmergingFriedmannII}.
\end{align}
Notice that from a macroscopic standpoint
Eqs.~(\ref{eq:EmergingFriedmannI}),~(\ref{eq:EmergingFriedmannII})
give the evolution of the emergent spacetime, with $\rho$ playing the
r{\^o}le of an auxiliary field (the modulus of $\sigma$) whose
dynamics is determined by the effective action in
Eq.~(\ref{eq:EffectiveAction}).

In this work we consider an effective potential of the following form
\be\label{eq:Potential}
\mathcal{V}(\sigma)=B|\sigma(\phi)|^2+\frac{2}{n}w|\sigma|^n+\frac{2}{n^{\prime}}w^{\prime}|\sigma|^{n^{\prime}},
\ee where we can assume $n^{\prime}>n$ without loss of generality. The
terms in the effective potential can be similarly motivated as in
Ref.~\cite{GFCEmergentFriedmann}. The interaction terms appearing in
GFT actions are usually defined in such a way that the perturbative
expansion of the GFT partition function reproduces that of spin foam
models. Specifically, spin foam models for $4d$ quantum gravity are
mostly based on interaction terms of power $5$, called simplicial. In
the case that the GFT field is endowed by a particular tensorial
transformation property, other classes of models can be obtained whose
interaction terms, called tensorial, are based on even powers of the
modulus of the field. In this light, the particular type of
interactions considered here can be understood as mimicking such types
of interactions, which is the reason why we will refer to them as
pseudosimplicial and pseudotensorial, respectively. In the following
we will study their phenomenological consequences, and show how
interesting physical effects are determined as a result of the
interplay between two interactions of this type. The integer-valued
powers $n$, $n^{\prime}$ in the interactions will be kept unspecified
throughout the article, thus making our analysis retain its full
generality. The particular values motivated by the above discussion
can be retrieved as particular cases. In the following we will show
how different ranges for such powers lead to phenomenologically
interesting features of the model, most notably concerning an early
era of accelerated expansion in Section~\ref{sec:MultiCriticalCase}.

Since $\mathcal{V}(\sigma)$ has to be bounded from below, we require
$w^{\prime}>0$. The equation of motion of the field $\sigma$ obtained
from Eqs. (\ref{eq:EffectiveAction}),~(\ref{eq:Potential}) is \be
-A\partial_{\phi}^2\sigma+B\sigma+w|\sigma|^{n-2}\sigma+w^{\prime}|\sigma|^{n^{\prime}-2}\sigma=0.
\ee Writing the complex field $\sigma$ in polar form (as in Eq.~(\ref{eq:PolarForm}), omitting indices)
$\sigma=\rho\;\e^{i\theta}$ one finds
(Ref.~\cite{GFCEmergentFriedmann}) that the equation of motion for the
angular component leads to the conservation law
\be\label{eq:ConservationOfQ}
\partial_{\phi}Q=0,\;\mbox{with}\;Q\equiv\rho^2
\partial_{\phi}{\theta}, \ee while the radial component satisfies a
second order ODE \be\label{eq:RadialEquation}
\partial^2_{\phi}\rho-\frac{Q^2}{\rho^3}-\frac{B}{A}\rho-\frac{w}{A}\rho^{n-1}-\frac{w^{\prime}}{A}\rho^{n^{\prime}-1}=0.
\ee The conserved charge $Q$ is proportional to the momentum of the
scalar field $\pi_{\phi}=\hbar Q$ \cite{GFCEmergentFriedmann}. One
immediately observes, that for large values of $\rho$ the term
$\rho^{n^{\prime}-1}$ becomes dominant.  In order to ensure that
Eq.~(\ref{eq:RadialEquation}) does not lead to drastic departures from
standard cosmology at late times (Eq.~(\ref{eq:EmergingFriedmannI})),
the coefficient of such term has to be positive
\be\label{eq:DefinitionMu} \mu\equiv-\frac{w^{\prime}}{A}>0, \ee which
implies, since $w^{\prime}>0$, that one must have $A<0$. In fact, the
opposite case $\mu<0$ would lead to an open cosmology expanding at a
faster than exponential rate, which relates to a Big Rip.
 Thus, considering $A<0$, compatibility with the free case (see
 Refs.~\cite{GFCEmergentFriedmann,FreeGFTpheno}) demands
 \be\label{eq:DefinitionMassSquared} m^2\equiv\frac{B}{A}>0, \ee which
 in turn implies $B<0$. The sign of $w$ is a priori not constrained,
 which leaves a considerable freedom in the model. Given the signs of
 the parameters $B$ and $w^{\prime}$, the potential in
 Eq.~(\ref{eq:Potential}) can be related to models with spontaneous
 symmetry breaking in Statistical Mechanics and Quantum Field
 Theory. The sub-leading term in the potential plays an important
 r{\^o}le in determining an inflationary-like era, as shown below in
 Section~\ref{sec:MultiCriticalCase}.

The connection to the theory of critical phenomena is to be expected
from the conjecture that GFT condensates arise through a phase
transition from a non-geometric to a geometric
phase~\cite{GFTGeometrogenesis}, which could be a possible realization
of the geometrogenesis scenario~\cite{Geometrogenesis}. Despite the
lack of a detailed theory near criticality and the fact that the
occurrence of the aforementioned phase transition is still a
conjecture, there are nonetheless encouraging results coming from the
analysis (both perturbative and non-perturbative) of the
Renormalization Group flow, which shows the existence of IR fixed
points in certain models, see Refs.~\cite{TGFTFRG,GFTRGReview}. On
this ground we will adopt, as a working hypothesis, the formation of a
condensate as a result of the phase transition, as in
Refs.~\cite{GFC,GFCOthers,GFCCalcagni,GFCEmergentFriedmann,GFClowspin,GFCReview}.

\begin{figure}
\includegraphics[width=\columnwidth]{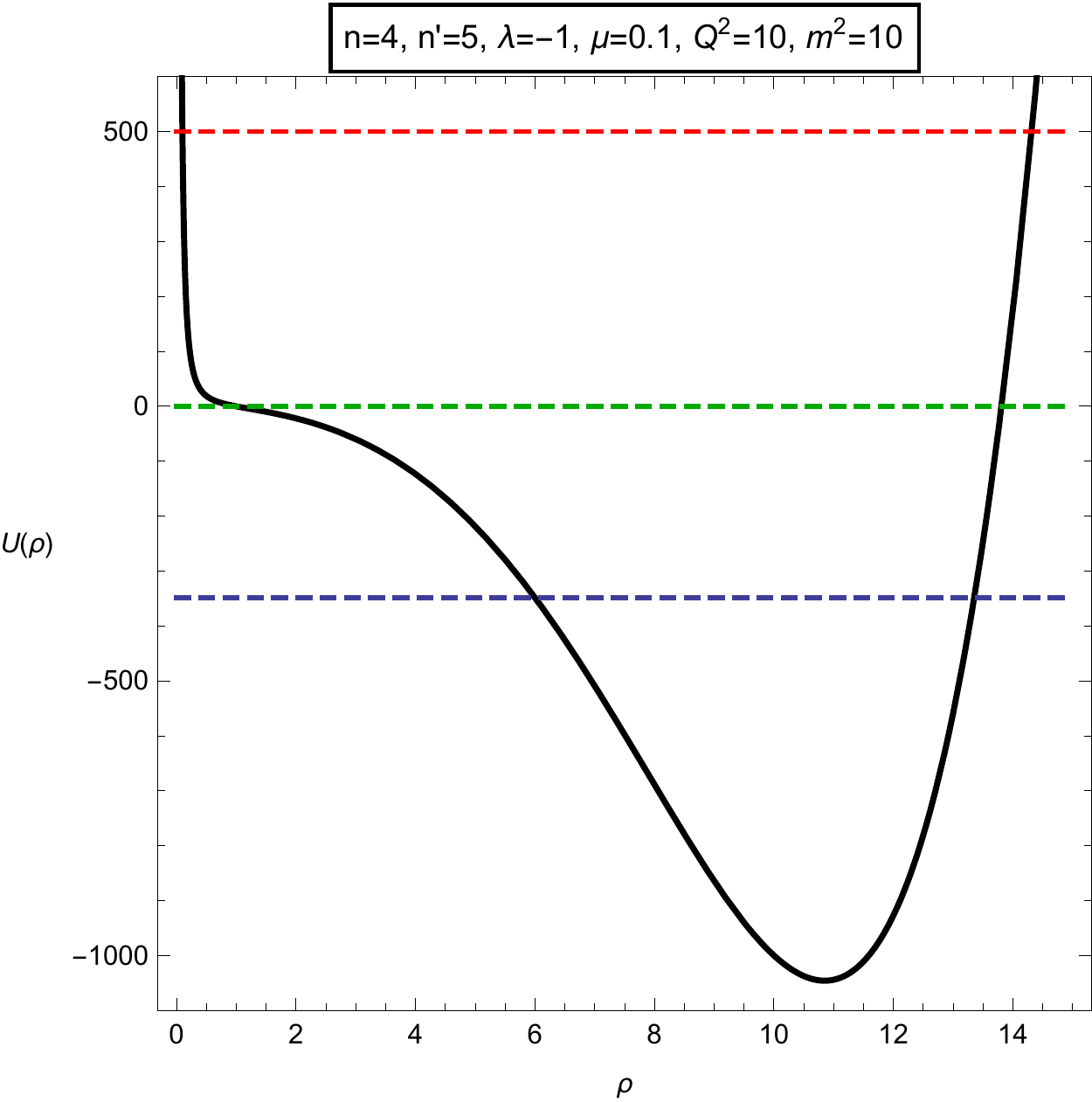}
\caption{Plot of the potential $U(\rho)$
  (Eq.~(\ref{eq:PointParticlePotentialU}) ) for the dynamical system
  described by Eq.~(\ref{eq:NewtonEquation}) and a particular choice
  of parameters. The three horizontal curves correspond to different
  values of the ``GFT energy'' $E$, in turn corresponding to different
  choices of initial conditions for $\rho$, $\rho^{\prime}$. The
  corresponding orbits in phase space are shown in
  Fig.~\ref{fig:PhasePortrait}. Recollapse is generic feature of the
  model and occurs for any values of the parameters, provided $\mu>0$
  and $Q\neq0$.}\label{fig:PotentialU}
\end{figure}

From Eqs.~(\ref{eq:DefinitionMu}),~(\ref{eq:DefinitionMassSquared})
and defining 
\be\label{eq:DefinitionLambda} \lambda\equiv-\frac{w}{A},
\ee 
we can rewrite Eq.~(\ref{eq:RadialEquation}) in the form
\be\label{eq:NewtonEquation} \partial^2_{\phi}\rho-m^2 \rho
-\frac{Q^2}{\rho^3}+\lambda\rho^{n-1}+\mu\rho^{n^{\prime}-1}=0~, \ee
that will be used throughout the rest of this article.  The above
equation has the form of the equation of motion of a classical point
particle with potential (see Fig.~\ref{fig:PotentialU})
\be\label{eq:PointParticlePotentialU}
U(\rho)=-\frac{1}{2}m^2\rho^2+\frac{Q^2}{2\rho^2}+\frac{\lambda}{n}\rho^n+\frac{\mu}{n^{\prime}}\rho^{n^{\prime}}.
\ee 
Equation~(\ref{eq:NewtonEquation}) leads to another conserved
quantity, $E$, defined as 
\be\label{eq:Energy}
E=\frac{1}{2}(\partial_{\phi}\rho)^2+U(\rho), \ee 
which is referred to
as ``GFT energy''~\cite{GFCEmergentFriedmann,FreeGFTpheno}. Its
physical meaning, from a fundamental point of view, is yet to be
clarified.

\begin{figure}
\includegraphics[width=\columnwidth]{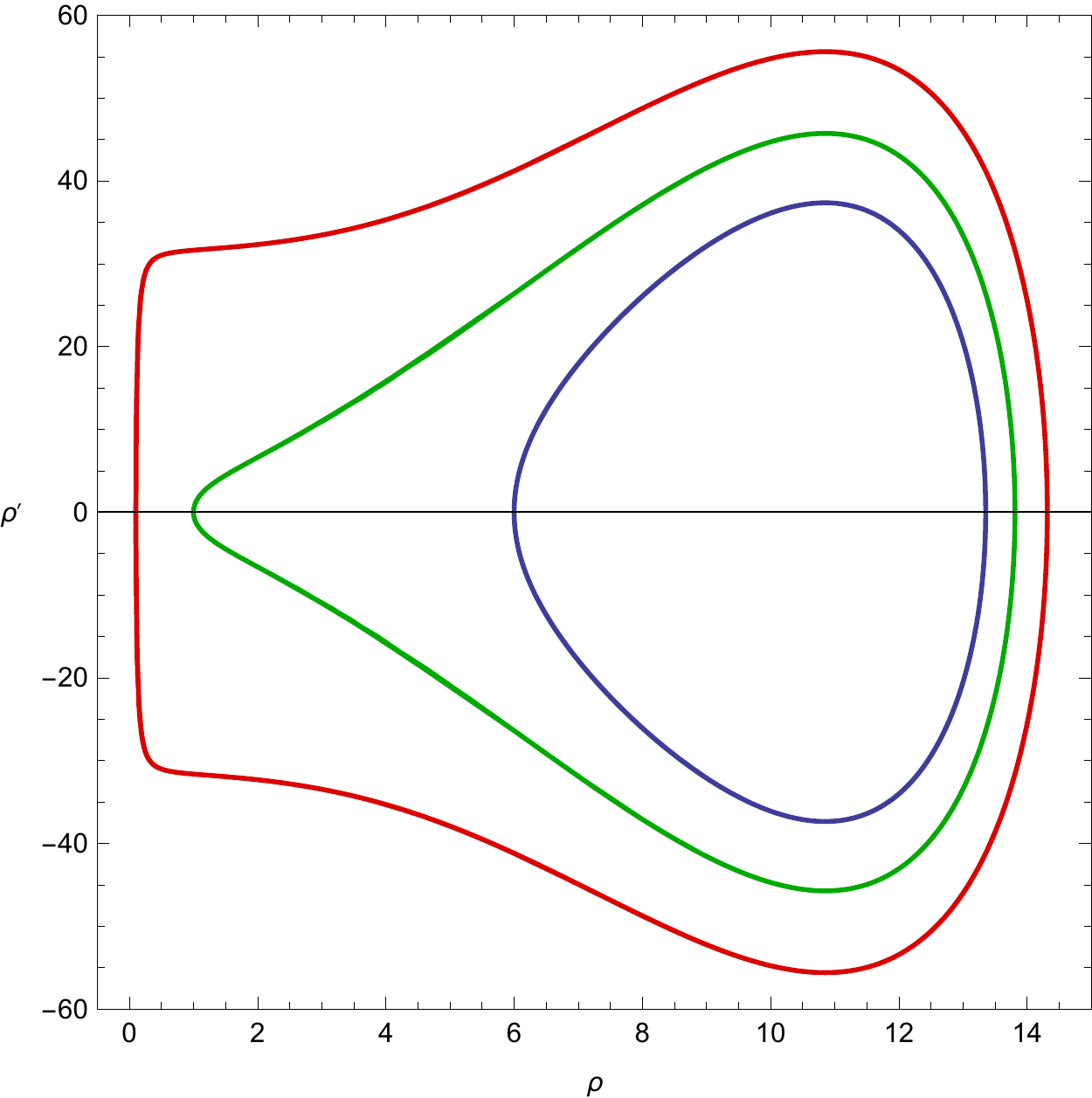}
\caption{Phase portrait of the dynamical system given by
  Eq.~(\ref{eq:NewtonEquation}). Orbits have energy given by the
  corresponding color lines as Fig.~\ref{fig:PotentialU}. Orbits are
  periodic and describe oscillations around the stable equilibrium
  point (\emph{center fixed point}) given by the absolute minimum of
  the potential $U(\rho)$. This is a general feature of the model
  which does not depend on the particular choice of parameters,
  provided Eq.~(\ref{eq:DefinitionMu}) is
  satisfied.}\label{fig:PhasePortrait}
\end{figure}

\begin{figure}
\includegraphics[width=\columnwidth]{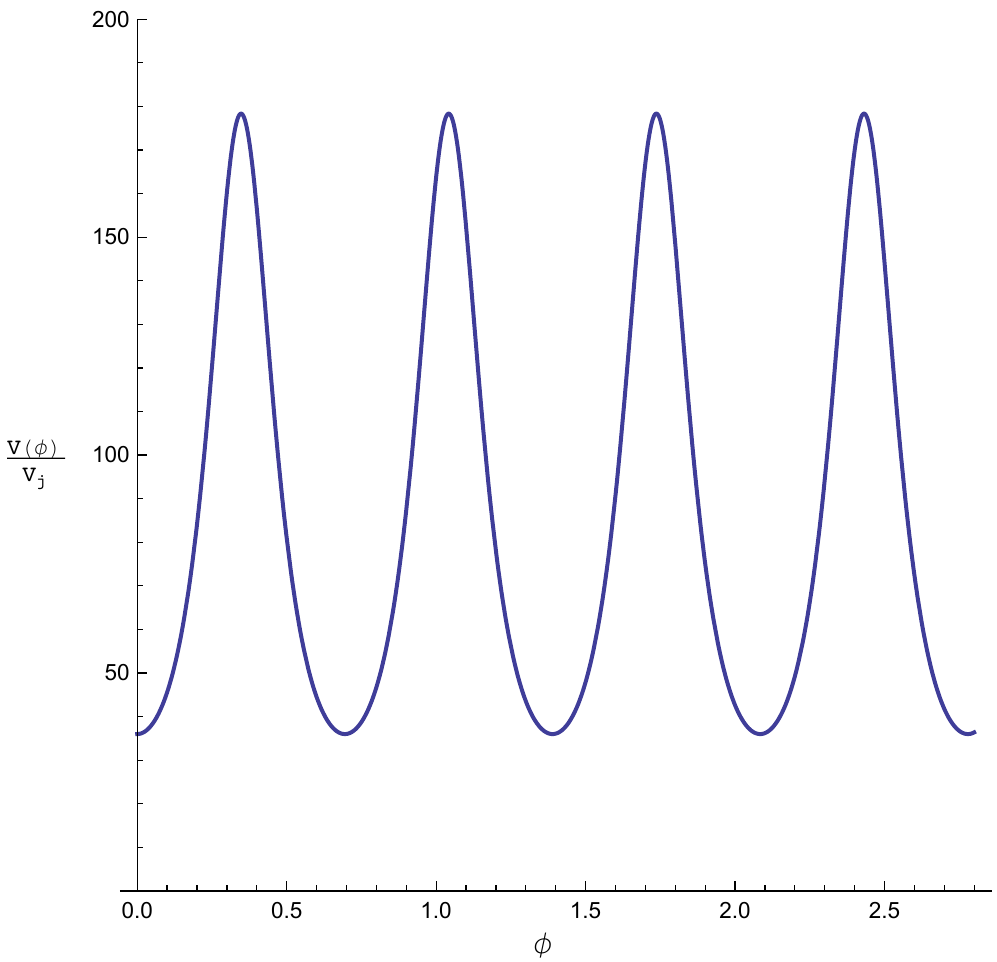}
\caption{Plot of the volume of the Universe as a function of
  relational time $\phi$ (in arbitrary units), corresponding to the
  blue orbit in Fig.~\ref{fig:PhasePortrait}. As a generic feature of
  the interacting model the Universe undergoes a cyclic evolution and
  its volume has a positive minimum, corresponding to a
  bounce.}\label{fig:Volume}
\end{figure}

\section{Recollapsing Universe}\label{sec:Recollapse}

Some properties of the solutions of the model and its consequences for
cosmology can already be drawn by means of a qualitative analysis of
the solutions of the second order ODE in
Eq.~(\ref{eq:PointParticlePotentialU}). In fact, solutions are
confined to the positive half-line $\rho>0$, given the infinite
potential barrier at $\rho=0$ for $Q\neq 0$. Moreover, since $\mu>0$
the potential in Eq.~(\ref{eq:PointParticlePotentialU}) approaches
infinity as $\rho$ takes arbitrarily large values. Therefore, provided
that we fix the ``GFT energy'' $E$ at a value which is larger than
both the absolute minimum and that of (possible) local maxima of the
potential $U(\rho)$, the solutions of Eq.~(\ref{eq:NewtonEquation})
turn out to be cyclic motions (see Figs.~\ref{fig:PotentialU},
\ref{fig:PhasePortrait}) describing oscillations around a stable
equilibrium point. These, in turn, correspond via
Eq.~(\ref{eq:VolumeDefinition}) to cyclic solutions for the dynamics
of the Universe Eqs.~(\ref{eq:EmergingFriedmannI}),
(\ref{eq:EmergingFriedmannII}) (see Fig.~\ref{fig:Volume}).

It is interesting to compare this result with what is known in the
case where interactions are
disregarded~\cite{GFCEmergentFriedmann,FreeGFTpheno}. In that case one
has that the Universe expands indefinitely and in the limit
$\phi\to\infty$ its dynamics follows the ordinary Friedmann equation
for a flat Universe, filled with a massless and minimally coupled
scalar field. Therefore, we see that the given interactions in the GFT
model induce a recollapse of the Universe, corresponding to the
turning point of the motion of $\rho$, as seen in
Fig.~\ref{fig:PhasePortrait}.

It is well-known in the classical theory that such a recollapse
follows as a simple consequence of the closed topology of 3-space. In
the GFC framework instead, the topology of space(time) is not fixed at
the outset, but should rather be reconstructed from the behavior of
the system in the macroscopic limit. In other words, the simple
condensate ansatz used here does not provide any information about the
topology of spatial sections of the emergent spacetime which, as it is
well known, play an important r{\^o}le in the dynamics of classical
cosmological models. Any topological information must therefore come
from additional input. A possible strategy one could follow is to work
with generalized condensates encoding such
information~\cite{GFTOperators}. Here, instead, we propose that the
closedness of the reconstructed space need not be encoded in the
condensate ansatz as an input, but is rather determined by the
dynamics as a consequence of the GFT interactions. Hence, allowing
only interactions that are compatible with reproducing a given spatial
topology, one may recover the classical correspondence between closed
spatial topology and having a finitely expanding Universe.

\section{Geometric inflation}\label{sec:GeometricInflation}

Cosmology obtained from GFT displays a number of interesting features
concerning the initial stage of the evolution of the Universe, which
mark a drastic departure from the standard FLRW cosmologies. In
particular, the initial big-bang singularity is replaced by a regular
bounce (see Ref.~\cite{GFCEmergentFriedmann,FreeGFTpheno}), followed
by an era of accelerated expansion \cite{FreeGFTpheno}. Similar
results were also obtained in the early LQC
literature, see \emph{e.g.}
Refs.~\cite{BojowaldBounce},~\cite{Bojowald}.  However, it is not
obvious \emph{a priori} that they must hold for GFT as
well. Nonetheless, it is remarkable that these two different
approaches yield qualitatively similar results for the dynamics of the
Universe near the classical singularity, even though this fact by
itself does not necessarily point at a deeper connection between the
two.

In the model considered in this paper, our results have a purely
quantum geometric origin and do not rely on the assumption of a
specific potential for the minimally coupled scalar
field\footnote{This is also the case in LQC, see
    Ref.~\cite{Bojowald}. However, the number of e-folds computed in
    that framework turns out to be too small in order to supplant
    inflation~\cite{BojowaldNotEnoughEFolds}.}, which is taken to be
massless and introduced for the sole purpose of having a relational
clock. This is quite unlike inflation, which instead heavily relies on
the choice of the potential and initial conditions for the inflaton in
order to predict an era of accelerated expansion with the desired
properties.

In this Section we investigate under which conditions on the
interaction potential of the GFT model it is possible to obtain an
epoch of accelerated expansion that could last long enough, so as to
account for the minimum number of e-folds required by standard
arguments. The number of e-folds is given by
\be\label{eq:DefinitionNumberE-FoldsVolume}
N=\frac{1}{3}\log\left(\frac{V_\mathrm{end}}{V_\mathrm{bounce}}\right),
\ee where $V_\mathrm{bounce}$ is the volume of the Universe at the
bounce and $V_\mathrm{end}$ is its value at the end of the era of
accelerated expansion. A necessary condition for it to be called an
inflationary era is that the number of e-folds must be large enough,
namely $N\gtrsim 60$.

Using Eq.~(\ref{eq:VolumeDefinition}) we rewrite
Eq.~(\ref{eq:DefinitionNumberE-FoldsVolume}) as
\be\label{eq:DefinitionNumberE-FoldsVolumeRho}
N=\frac{2}{3}\log\left(\frac{\rho_\mathrm{end}}{\rho_\mathrm{bounce}}\right),
\ee with an obvious understanding of the notation. This formula is
particularly useful since it allows us to derive the number of e-folds
only by looking at the dynamics of $\rho$.

Since there is no notion of proper time, a sensible definition of
acceleration can only be given in relational terms. In particular, we
seek a definition that agrees with the standard one given in ordinary
cosmology. As in Ref.~\cite{FreeGFTpheno} we can therefore define the
acceleration as \be\label{eq:DefinitionAcceleration}
\mathfrak{a}(\rho)\equiv\frac{\partial^2_{\phi}V}{V}-\frac{5}{3}\left(\frac{\partial_{\phi}V}{V}\right)^2.
\ee Hence, from Eqs.~(\ref{eq:EmergingFriedmannI}),
(\ref{eq:EmergingFriedmannII}) one gets the following expression for
the acceleration $\mathfrak{a}$ as a function of $\rho$ for a generic
potential \be
\mathfrak{a}(\rho)=-\frac{2}{\rho^2}\left\{\partial_{\phi}U(\rho )
\rho +\frac{14}{3}\left[E-U(\rho )\right]\right\}.  \ee Using
Eq.~(\ref{eq:PointParticlePotentialU}) one finally has for our model
\be\label{eq:AccelerationVsRho}
\begin{split}
\mathfrak{a}(\rho)&=-\frac{2}{\rho^2}\left[\frac{14}{3}E+\left(1  -\frac{14
   }{3 n^{\prime}}\right) \mu\rho^{n^{\prime}}+\frac{4 m^2 \rho ^2}{3}\right.\\
&+\left.\left(1 -\frac{14  }{3n}\right)\lambda \rho
   ^n-\frac{10 Q^2}{3 \rho ^2}\right].
\end{split}
\ee 
Therefore, the sign of the acceleration is opposite to that of the
polynomial 
\be s(\rho)=P(\rho)+\left(3 -\frac{14 }{n}\right)\lambda
\rho ^{n+2}+\left(3 -\frac{14 }{ n^{\prime}}\right)
\mu\rho^{n^{\prime}+2}, \ee 
where we defined 
\be P(\rho)=4 m^2 \rho
^4+14 E\rho^2-10 Q^2.  \ee In the following we will study in detail
the properties of the era of accelerated expansion. The free case will
be discussed in Section~\ref{sec:FreeCase}, whereas the r{\^o}le of
interactions in allowing for an inflationary-like era will be the
subject of Section~\ref{sec:MultiCriticalCase}.

\subsection{The non-interacting case}\label{sec:FreeCase}

In this case the acceleration is given by 
\be
\mathfrak{a}(\rho)=-\frac{2}{3\rho^4}P(\rho).  \ee
The bounce occurs
when $\rho$ reaches its minimum value, \emph{i.e.} when $U(\rho)=E$,
leading to 
\be\label{eq:RhoMinimumFreeCase}
\rho_\mathrm{bounce}^2=\frac{1}{m^2}\left(\sqrt{E^2+m^2 Q^2}-E\right).
\ee 
A straightforward calculation shows that
$\mathfrak{a}(\rho_\mathrm{bounce})>0$ as expected. The era of
accelerated expansion ends when $P(\rho)$ vanishes, which happens at a
point $\rho_{\star}>\rho_\mathrm{bounce}$, which is given by
\be\label{eq:RhoEndFreeCase}
\rho_{\star}=\frac{1}{4m^2}\left(\sqrt{49E^2+40m^2 Q^2}-7E\right).
\ee We can then use Eqs.~(\ref{eq:DefinitionNumberE-FoldsVolumeRho}),
(\ref{eq:RhoMinimumFreeCase}), (\ref{eq:RhoEndFreeCase}) to determine the
energy $E$ as a function of the number of e-folds $N$. Reality of $E$
thus leads to the following bounds on $N$ \be
\frac{1}{3}\log\left(\frac{10}{7}\right)\leq N\leq
\frac{1}{3}\log\left(\frac{7}{4}\right), \ee that is \be 0.119\lesssim
N\lesssim 0.186.  \ee Such tight bounds, holding for all values of the
parameters $m^2$ and $Q^2$, rule out the free case as a candidate to
replace the standard inflationary scenario in cosmology.

\subsection{The interacting case: the multicritical model}\label{sec:MultiCriticalCase}

In this subsection we investigate the consequences of interactions for
the evolution of the Universe. In particular, we show how the
interplay between the two interaction terms in the effective potential
(Eq.~\ref{eq:Potential}) makes it possible to have an early epoch of
accelerated expansion, which lasts as long as in inflationary models.
Before studying their effect, we want to discuss how the occurrence of
such interaction terms could be motivated from the GFT perspective. In
principle, one could have infinitely many interaction terms given by
some power of the GFT field. However, only a finite number of them
will be of relevance at a specific scale, as dictated by the behaviour
of the fundamental theory under the Renormalization Group (RG) flow.

In a continuum and large scale limit new terms in the action could be
generated, whereas others might become irrelevant. In this sense, one
might speculate that, e.g., in addition to the five-valent simplicial
interaction term the effective potential includes another term which
becomes relevant on a larger scale. Ultimately, rigorous RG arguments
will of course have the decisive word regarding the possibility to
obtain such terms from the fundamental theory. Nevertheless, by
studying the phenomenological features of such potentials and
extracting physical consequences from the corresponding cosmological
solutions, we aim at clarifying the map between the fundamental
microscopic and effective macroscopic dynamics of the theory. At the
same time, our results might help to shed some light onto the subtle
issue of the physical meaning of such interaction terms.

Hereafter we assume the hierarchy $\mu\ll |\lambda|$, since otherwise
an inflationary era cannot be easily accommodated.This means that the
higher order term in the interaction potential $\mathcal{V}(\sigma)$
becomes relevant only for very large values of the condensate field
$\sigma$, hence of the number of quanta representing the basic
building blocks of quantum spacetime. Consequently, the dynamics in
the immediate vicinity of the bounce is governed by the parameters of
the free theory and the sub-leading interaction term.

To begin with, let us start by fixing the value of the GFT energy. We
require the Universe to have a Planckian volume at the bounce. Since
the volume is given by Eq.~(\ref{eq:VolumeDefinition}), this is done
by imposing $\rho_\mathrm{bounce}=1$. Such a condition also fixes the
value of the GFT energy to \be E=U\left(\rho_\mathrm{bounce}=1\right).
\ee In fact, we demand that $\rho_\mathrm{bounce}$ is the minimal
value of $\rho$ which is compatible with the GFT energy $E$ available
to the system. Hence, we also have the condition
$\partial_{\rho}U\left(\rho_\mathrm{bounce}=1\right)\leq0$. Notice
that this is trivially satisfied in the free case. In the interacting
case (holding the hierarchy $\mu\ll |\lambda|$) one can therefore use
it to obtain a bound on $\lambda$ \be\label{first constraint on
  lambda} \lambda\leq m^2+Q^2.  \ee It is convenient for our purposes
and in order to carry over our analysis in full generality, to
introduce the definitions
\begin{align}
\alpha&\equiv\left(3-\frac{14}{n}\right)\lambda\label{eq:DefinitionAlpha},\\
\beta&\equiv\left(3-\frac{14}{n^{\prime}}\right)\mu\label{eq:DefinitionBeta}.
\end{align}
The acceleration Eq.~(\ref{eq:AccelerationVsRho}) can thus be written
as 
\be\label{eq:AccelerationAlphaBeta}
\mathfrak{a}(\rho)=-\frac{2}{\rho^4}\left[P(\rho)+\alpha
  \rho^{n+2}+\beta\rho^{n^{\prime}+2}\right].  \ee 
As pointed out
before, $\mathfrak{a}>0$ has to hold at the bounce. The first thing to
be observed is that $\alpha<0$ is a necessary condition in order to
have enough e-folds. In fact, if this were not the case, the bracket
in Eq.~(\ref{eq:AccelerationAlphaBeta}) would have a zero at a point
$\rho_\mathrm{end}<\rho_{\star}$ (cf. Eq.~(\ref{eq:RhoEndFreeCase})),
thus leading to a number of e-folds which is even smaller than the
corresponding one in the free case. Furthermore, it is possible to
constrain the value of $\mu$ in a way that leads both to the
aforementioned hierarchy and to the right value for $N$, which we
consider as fixed at the outset. In order to do so, we solve
Eq.~(\ref{eq:DefinitionNumberE-FoldsVolumeRho})
w.r.t. $\rho_\mathrm{end}$, having fixed the bounce at
$\rho_\mathrm{bounce}=1$ \be\label{eq:EndBounceN}
\rho_\mathrm{end}=\rho_\mathrm{bounce}~\mbox{e}^{\frac{3}{2}N}.  \ee
The end of inflation occurs when the polynomial in the bracket in
Eq.~(\ref{eq:AccelerationAlphaBeta}) has a zero. Since
$\rho_\mathrm{end}\gg1$, it is legitimate to determine this zero by
taking into account only the two highest powers in the polynomial,
with respect to which all of the other terms are negligible. We
therefore have \be
\alpha\rho_\mathrm{end}^{n+2}+\beta\rho_\mathrm{end}^{n^{\prime}+2}\approx0,
\ee which, using Eq.~(\ref{eq:EndBounceN}), leads to
\be\label{eq:AlphaBetaN} \beta=-\alpha \; \e^{-\frac{3}{2} N
  (n^{\prime}-n)}.  \ee The last equation is consistent with the
hierarchy $\mu\ll |\lambda|$ and actually fixes the value of $\mu$
once $\lambda$, $n$, $n^{\prime}$ and $N$ are assigned. Furthermore
one has $\beta>0$ which, together with
Eqs.~(\ref{eq:DefinitionMu}),~(\ref{eq:DefinitionBeta}), implies
$n^{\prime}>\frac{14}{3}$. Importantly, this means that $n^{\prime}=
5$ is the lowest possible integer compatible with an inflationary-like
era. This particular value is also interesting in another respect
since in GFT typically only specific combinatorially non-local
interactions minimally of such a power allow for an interpretation in
terms of simplicial quantum gravity \cite{GFT,GFT4All}.

Our considerations so far leave open two possibilities, viz.:
 \begin{itemize}
 \item $\lambda<0$ and $n\geq5$ ($n^{\prime}>n$), which in the case of $n=5$ could correspond to the just mentioned simplicial interaction term and the higher order $n'$-term could possibly be generated in the continuum and large scale limit of the theory and becomes dominant for very large $\rho$. For even $n'$ it mimics so-called tensorial interactions.
\item $\lambda>0$ and $2<n<5$ ($n^{\prime}\geq5$), which for $n'=5$ could allow a connection to simplicial quantum gravity and would remain dominant for large $\rho$ over the $n$-term, which in the case $n=4$ is reminiscent of an interaction of tensorial type.
  \end{itemize}

\begin{figure}
\includegraphics[width=\columnwidth]{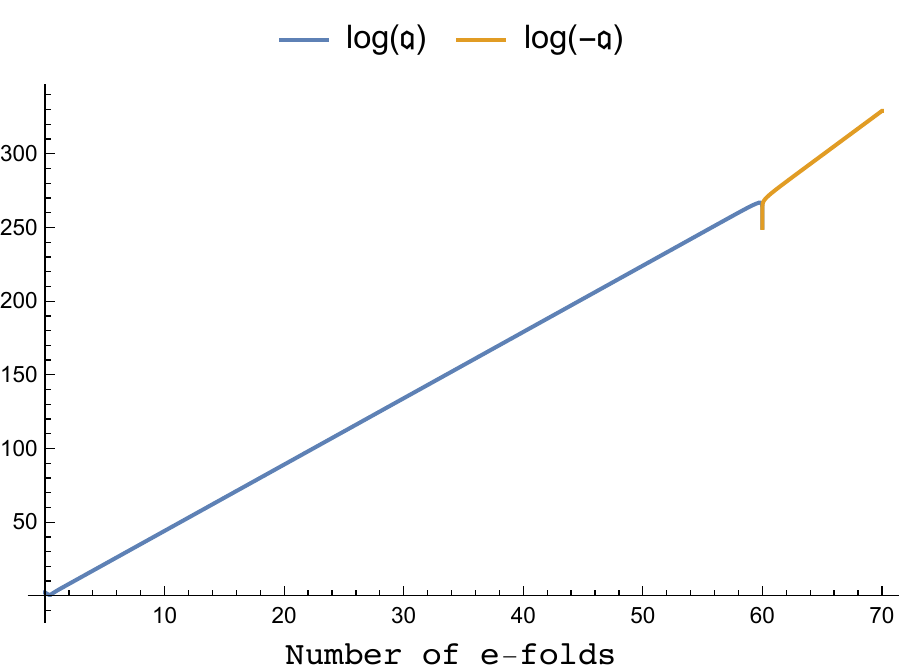}
\caption{Inflationary era supported by GFT interactions in the
  multi-critical model. The blue (orange) curve represents the graph
  of the logarithm of the acceleration (minus the acceleration) as a
  function of the number of e-folds in the case $\lambda<0$. The plot
  refers to the particular choice of parameters $n=5$, $n^{\prime}=6$,
  $m=1$, $Q=1$, $\lambda=-3$. The value of $\mu$ is determined from
  Eq.~(\ref{eq:AlphaBetaN}) by requiring the number of e-folds to be
  $N=60$. There is a logarithmic singularity at $N\simeq 60$, marking
  the end of the accelerated
  expansion. Figs.~\ref{fig:eFoldsBounce},~\ref{fig:eFoldsEnd} show
  the behavior of the acceleration close to the bounce and at the end
  of inflation respectively.}\label{fig:eFolds}
\end{figure}

\begin{figure}
\includegraphics[width=\columnwidth]{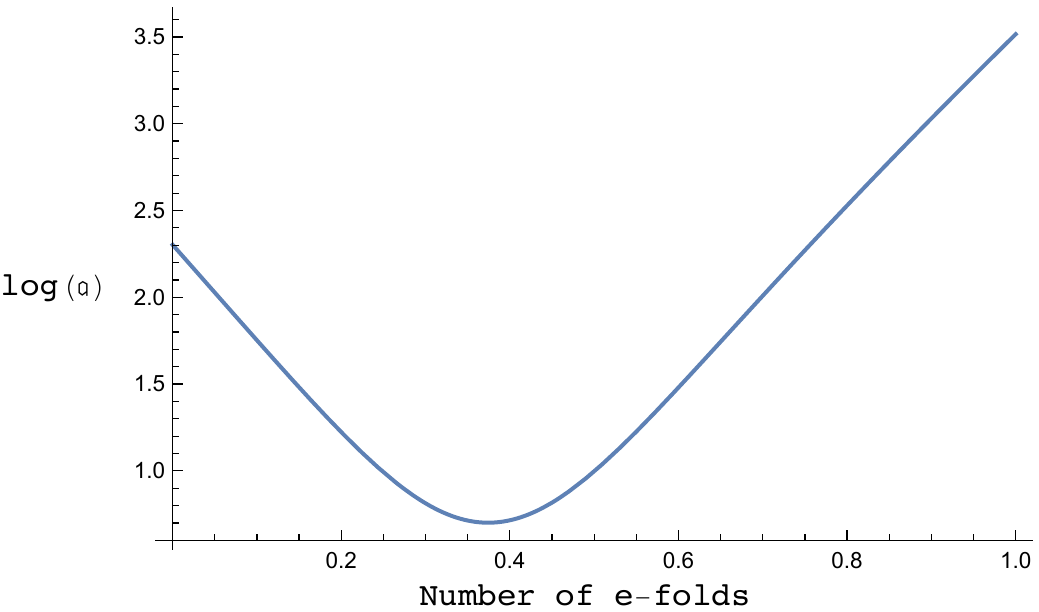}
\caption{}\label{fig:eFoldsBounce}
\end{figure}

\begin{figure}
\includegraphics[width=\columnwidth]{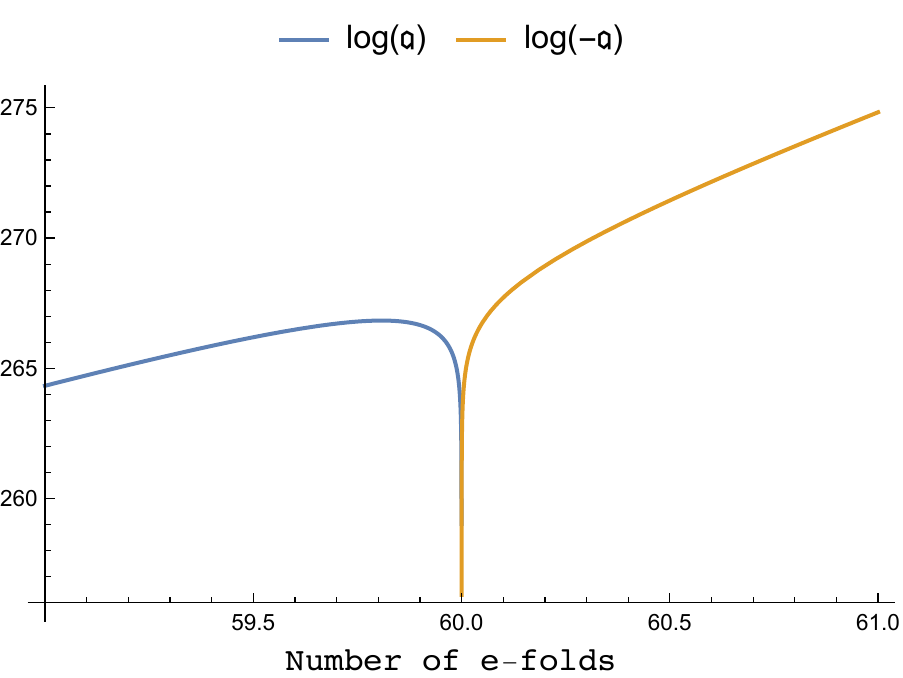}
\caption{}\label{fig:eFoldsEnd}
\end{figure}

However, this is not yet enough in order to guarantee an
inflation-like era. In fact we have to make sure that there is no
intermediate stage of deceleration occurring between the bounce at
$\rho_b=1$ and $\rho_\mathrm{end}$, \emph{i.e.}, that
$\mathfrak{a}(\rho)$ stays positive in the interval between these two
points. In other words we want to make sure that $\rho_\mathrm{end}$
is \emph{the only zero} of the acceleration lying to the right of
$\rho_b$. In fact $\mathfrak{a}(\rho)$ starts positive at the bounce
and has a minimum when $P(\rho)$ becomes of the same order of
magnitude of the term containing the power $\rho^{n+2}$ (see
Eq.~(\ref{eq:AccelerationAlphaBeta})).  Thus we see that we have to
require that the local minimum of $\mathfrak{a}(\rho)$ (\emph{i.e.}
the maximum of the poynomial in brackets in
Eq.~(\ref{eq:AccelerationAlphaBeta})) is positive (resp. negative). As
$\rho$ increases further, the acceleration increases again until it
reaches a maximum when the contribution coming from the term
containing $\rho^{n^{\prime}+2}$ becomes of the same order of
magnitude of the other terms. Thereafter the acceleration turns into a
decreasing function all the way until $\rho\to+\infty$ and therefore
has a unique zero. Positivity of the local minimum of
$\mathfrak{a}(\rho)$ translates into a further constraint on
parameters space. By direct inspection, it is possible to see that the
latter case listed above does not satisfy such condition for any value
of the parameters of the model. Therefore we conclude that
\emph{$\lambda$ must be negative} if the acceleration is to keep the
same sign throughout the inflationary era. The evolution of the
acceleration as a function of relational time $\phi$ is shown in
Figs.~\ref{fig:eFolds},~\ref{fig:eFoldsBounce},~\ref{fig:eFoldsEnd}
for some specific choice of the parameters. It is worthwhile stressing
that the behaviour of the model in the case $\lambda<0$ is
nevertheless generic and therefore does not rely on the specific
choice of parameters. Furthermore, by adjusting the value of $N$ and
the other parameters in Eq.~(\ref{eq:AlphaBetaN}), it is possible to
achieve any desirable value of e-folds during inflation.

All we said in this Section applies to the multi-critical model with
the effective potential Eq.~(\ref{eq:Potential}) but \emph{does not
  hold} in a model with only one interaction term. In fact in that
case it is not possible to prevent the occurrence of an intermediate
era of deceleration between $\rho_{b}$ and $\rho_\mathrm{end}$, the
latter giving the scale at which the higher order interaction term
becomes relevant.

One last remark is in order: inflation was shown to be a feature of
multicritical GFT models but only at the price of a fine-tuning in the
value of the parameter $\mu$ (see Eq.~(\ref{eq:AlphaBetaN})).

\section{Interactions and the final fate of the Universe}\label{sec:EffectiveFriedmannEffFluids}

It is possible to recast the dynamical equations for the volume of the
Universe in a form that bears a closer resemblance to the standard
Friedmann equation, as shown in Ref.~\cite{FreeGFTpheno}. In fact, the
Hubble expansion rate can be expressed as (see the appendix for more
details) 
\be\label{eq:HubbleRate}
H=\frac{1}{3}\frac{\partial_{\phi}V}{V^2}\pi_{\phi}.  \ee 
From
Eq.~(\ref{eq:EmergingFriedmannI}) and the proportionality between the
momentum of the scalar field and $Q$ we have
\be\label{eq:FriedmannHubbleRate} H^2=\frac{4}{9}\frac{\hbar^2 Q^2}{
  V^2}\left(\frac{\partial_{\phi}\rho}{\rho}\right)^2.  \ee 
The term
in bracket can thus be interpreted as a dynamical effective
gravitational constant as in Ref.~\cite{FreeGFTpheno}.  Alternatively,
using
Eqs.~(\ref{eq:VolumeDefinition}),~(\ref{eq:Energy}),~(\ref{eq:PointParticlePotentialU}),
the last equation Eq.~(\ref{eq:FriedmannHubbleRate}) becomes
(considering the case with only one interaction term, namely
$\lambda=0$) 
\be\label{eq:FriedmannEffectiveFluids} H^2=\frac{8
  \hbar^2 Q^2}{9}
\left[\frac{\varepsilon_m}{V^2}+\frac{\varepsilon_E}{V^3}+\frac{\varepsilon_Q}{V^4}+\frac{\varepsilon_\mu}{V^{3-n^{\prime}/2}}\right],
\ee 
where we defined
\begin{align}
\varepsilon_E&=V_j E,\\
\varepsilon_m&=\frac{m^2}{2},\\
\varepsilon_Q&=-\frac{Q^2}{2}V_j^2,\\
\varepsilon_{\mu}&=-\frac{\mu}{n^{\prime}}V_j^{1-n^{\prime}/2}.
\end{align}
The exponents of the denominators in
Eq.~(\ref{eq:FriedmannEffectiveFluids}) can be related to the $w$
coefficients in the equation of state $p=w \varepsilon$ of some
effective fluids, with energy density $\varepsilon$ and pressure
$p$. Each term scales with the volume as $\propto V^{-(w+1)}$.

It is worth pointing out that Eq.~(\ref{eq:FriedmannEffectiveFluids})
makes clearer the correspondence with the framework of ekpyrotic
models\footnote{We are thankful to Martin Bojowald for this
  observation.}, where one has the gravitational field coupled to
matter fields with $w>1$. Such models have been advocated as a
possible alternative to inflation, see Ref.~(\cite{Ekpyrosis}). We
observe that at early times (\emph{i.e.} small volumes) the occurrence
of the bounce is determined by the negative sign of $\varepsilon_{Q}$,
which is also the term corresponding to the highest $w$. However,
while this is sufficient to prevent the classical singularity, it is
not enough to guarantee that the minimum number of e-folds is reached
at the end of the accelerated expansion, as shown in
Section~\ref{sec:FreeCase}. In fact, the r{\^o}le of interactions is
crucial in that respect, as our analysis in
Section~\ref{sec:MultiCriticalCase} has shown.

In the rest of this Section we focus instead on the consequences of
having interactions in the GFT model for the evolution of the Universe
at late times. As we have already seen in Section
\ref{sec:Recollapse}, a positive $\mu$ entails a recollapsing
Universe. This should also be clear from
Eq.~(\ref{eq:FriedmannEffectiveFluids}). In particular, we notice that
the corresponding term in the equation is an increasing function of
the volume for $n^{\prime}>6$. This is quite an unusual feature for a
cosmological model, where all energy components (with the exception of
the cosmological constant) are diluted by the expansion of the
Universe. For $n^{\prime}=6$ one finds instead a cosmological constant
term. It is also possible to have the interactions reproduce the
classical curvature term $\propto \frac{\kappa}{V^{2/3}}$ by choosing
$n^{\prime}=\frac{14}{3}$, which is however not allowed if one
restricts to integer powers in the interactions \cite{GFT}.

Our analysis shows that only $\lambda<0$ leaves room for an era
accelerated expansion analogous to that of inflationary models. In
order for this to be possible, one must also have $n^{\prime}>n\geq
5$. Moreover, if one rules out phantom energy(\emph{i.e.} $w<-1$), there
is only one case which is allowed, namely $n=5$, $n^{\prime}=6$. Then
during inflation the Universe can be described as dominated by a fluid
with equation of state $w=-\frac{1}{2}$. After the end of inflation
its energy content also receives contribution from a negative
cosmological constant, which eventually leads to a recollapse. It is
remarkable that this particular case selects an interaction term which
is in principle compatible with the simplicial interactions which have
been extensively considered in the GFT approach. However, it must be
pointed out that the realization of the geometric inflation picture
imposes strong restrictions also on the type of interactions one can
consider, as well as on their relative strength.

\section{Summary and Outlook}\label{sec:Outlook}

We investigated the phenomenological consequences of interacting GFT
models for the dynamics of the early Universe, which in this framework
is seen as emerging from the collective behavior of ``quanta of
geometry''. The dynamical equations for the GFT condensate are
classical and are obtained from an effective action, whose form can be
motivated from the microscopic theory. The dynamics of the condensate
in turn determines a Friedmann-like dynamics for the geometric
observables corresponding to classical dynamical variables.

In this article we considered an effective potential including two
interaction terms besides the quadratic one, the latter being already
present in the free theory. An ambiguity in the kinetic term,
represented by the factor $A$, is fixed by requiring the expansion of
the Universe not to be faster than exponential at large volumes. A
general prediction of the model is the occurrence of a recollapse when
the higher order interaction term becomes codominant. Results that
have already been obtained in the free theory
(Refs.~\cite{GFCEmergentFriedmann,FreeGFTpheno}) survive in the
interacting case, in particular for what concerns the occurrence of a
bounce and an early epoch of accelerated expansion. The former result,
together with the recollapse induced by interactions, leads to cyclic
cosmologies. A more detailed analysis of the latter, instead, leads to
the conclusion that, in the free case, the era of accelerated
expansion does not last for a number of e-folds which is at least as
large as in inflationary models. This is instead made possible when
suitable interaction terms are taken into account, as in the
multi-critical model considered here. Indeed, we showed that one can
attain an arbitrary number of e-folds as the Universe accelerates
after the bounce. Furthermore, having an inflationary-like expansion
imposes a restriction on the class of viable models. In fact, this is
possible only for $\lambda<0$ and $5\leq n<n^{\prime}$ and when one
has the hierarchy $\mu\ll |\lambda|$. Reasonable phenomenological
arguments lead to select only the case $n=5$, $n^{\prime}=6$ as
physical. The two powers can be related, respectively, to simplicial
interactions, commonly considered in the GFT framework, and to a
negative cosmological constant. While the result is encouraging as a
first step towards a quantum geometric description of the inflationary
era, a few remarks are in order. In fact, it must be pointed out that
it comes at the price of a fine tuning in the coupling constant of the
higher order interaction term. From this point of view, it shares one
of the major difficulties of ordinary inflationary
models. Furthermore, an inflationary-like era does not seem to be a
generic property of GFT models, but in fact requires interactions of a
suitable form.

Future work must be devoted to incorporate other degrees of freedom in the description of the effective dynamics of the emergent spacetime. In fact their phenomenological signatures, in particular for what concerns the seeds for the growth of structures, are crucial in order to be able to give a definite answer to the problem of finding a valid alternative to the inflationary paradigm, which might come from quantum geometry. 

Cyclic cosmologies were obtained under fairly general assumptions on
the effective potential, namely its boundedness from
below. Furthermore, no input was given about the geometry of the
spatial sections of the reconstructed spacetime. Given the classical
correspondence existing between a finitely expanding Universe and
closed spatial topology, it is intriguing whether this purely
dynamical result bears any implications on the geometry of the
emergent spacetime. In particular it would be worth investigating the
existence of such a generalized correspondence by extending our
analysis studying, \emph{e.g.}, the implications of the peculiar
combinatorial structure of GFT interactions for the geometry of the
emergent spacetime and its dynamics. In fact, this further step would
be required, in order to properly take into account the effects of
anisotropies.

In this article we considered a model that can in principle exhibit
multicritical behaviour, which is reminiscent of analogous models
discussed in statistical field theory to model systems with a more
complex phase structure~\cite{SFTQFT}. Multicriticality has not been
considered in the GFT literature so far. Nevertheless it would be
worth studying the possible implications of such models also from a
fundamental GFT point of view and relate them to the geometrogenesis
scenario.

It is our hope that the interplay between determining
phenomenological constraints, as those obtained in this work, and a
fundamental approach involving, \emph{e.g.}, RG arguments, might help
to single out the correct microscopic theory (or even a family of such
theories). Further work must come from both directions in a common
effort to develop an appropriate framework for studying early Universe
cosmology, which correctly takes into account the quantum dynamics of
all the relevant degrees of freedom of the gravitational field.

We would like to draw the attention of the reader to another important aspect concerning the interpretation of the results obtained in this work, with respect to the physical significance and the r{\^o}le played by the operator counting the number of nodes in the spin network, whose expectation value is given by $\mathcal{N}=\sum_j\rho_j^2$. Looking back at Eq.~(\ref{eq:VolumeDefinitionGeneral}), one sees that in the case in which the system lies just in one representation with fixed $j$ (as supported by the findings in Ref.~\cite{GFClowspin}) this quantity is proportional to the total volume of the universe. It is therefore clear that only its relative variations are observable, and turn out to be proportional to the Hubble expansion rate.

As a closing remark, we would like to stress that the GFT condensate cosmology approach is not the only one which has been suggested in the literature to extract the cosmological sector of LQG from a covariant formulation of its dynamics. In fact, spin foam cosmology (SFC) \cite{SFC} makes use of the spin foam expansion \cite{SF} and is thus an expansion in terms of the number of degrees of freedom. Initially, SFC was studied using the simplest cellular decomposition of the $3$-sphere, given by the so-called dipole graph \cite{DipoleCosmologies}, and was later on extended to more general regular graphs \cite{SFCGenGraph}. Central to this approach is the assumption that the relevant physics is encoded within a fixed number of quanta of geometry, whereas in GFT condensate cosmology there is no a priori restriction on the number of quanta which - in principle - can be large. As suggested in \cite{GFC} and \cite{GFCReview}, a GFT condensate could also be constituted by means of dipoles or even more complicated building blocks and it would be important to repeat our analysis for those and compare the results to the ones obtained from SFC.

\appendix*
\section{}
Here we review how the effective equations that give the relational
evolution of the volume of the Universe,
Eqs.~(\ref{eq:EmergingFriedmannI}), (\ref{eq:EmergingFriedmannII}) can
be recast in a form that is closer to that of ordinary FLRW models. In
fact, in the framework considered in this article, spacetime is an
emergent concept. Hence, there is no natural notion of proper time and
evolution of physical observables is more appropriately defined in
terms of a matter clock, here represented by a massless scalar which
is minimally coupled to the gravitational field. It is nonetheless
worth showing the relation that exists between the two different
descriptions in the classical theory, where they are both well defined
and by considering the classical relation between the velocity of the
scalar field $\phi$ and its canonically conjugate momentum
$\pi_{\phi}$ \be\label{eq:Momentum} \pi_{\phi}=V\dot{\phi}, \ee
together with the relation between the proper volume and the scale
factor \be\label{eq:VolumeScaleFactor} V\propto a^3, \ee one can
express the Hubble expansion rate as
\be\label{eq:AppendixFirstDerVolume}
H=\frac{\dot{a}}{a}=\frac{1}{3}\frac{\dot{V}}{V}=\frac{1}{3}\pi_{\phi}\frac{\partial_{\phi}V}{V^2}.
\ee The last equation is the same as Eq.~(\ref{eq:HubbleRate}), which
leads to the modified Friedmann equation of
Eq.~(\ref{eq:FriedmannHubbleRate}). In a similar fashion, it is also
possible to get the Raychaudhuri equation for the acceleration. In
fact, taking two derivatives of Eq.~\ref{eq:VolumeScaleFactor} one is
lead to \be\label{eq:AppendixSecondDerScale}
\frac{\ddot{a}}{a}=\frac{1}{3}\left[\frac{\ddot{V}}{V}-\frac{2}{3}\left(\frac{\dot{V}}{V}\right)^2\right].
\ee Furthermore, using Eq.~(\ref{eq:Momentum}), one finds \be
\dot{V}=\partial_{\phi}V \dot{\phi}=\partial_{\phi}V
\frac{\pi_{\phi}}{V} \ee and \be\label{eq:AppendixSecondDerVolume}
\ddot{V}=\left(\frac{\pi_{\phi}}{V}\right)^2 \left[\partial^2_{\phi}V
  - \frac{\left(\partial_{\phi}V\right)^2}{V} \right].  \ee From
Eqs.~(\ref{eq:AppendixFirstDerVolume}),~(\ref{eq:AppendixSecondDerScale}),~(\ref{eq:AppendixSecondDerVolume})
one has \be
\frac{\ddot{a}}{a}=\frac{1}{3}\left(\frac{\pi_{\phi}}{V}\right)^2\left[\frac{\partial^2_{\phi}V}{V}-\frac{5}{3}\left(\frac{\partial_{\phi}V}{V}\right)^2\right].
\ee The last equation justifies the definition of the acceleration
given in Eq.~(\ref{eq:DefinitionAcceleration}).


\end{document}